\documentclass[epj,final]{svjour}

\usepackage{color}
\usepackage{graphicx}
\usepackage{dcolumn}
\usepackage{epsfig}
\usepackage{epstopdf}
\usepackage{bm}
\usepackage[Gray,squaren]{SIunits}
\graphicspath{{fig/}}
\usepackage{soul}

\newcommand{\snS}{{\rm S}}
\newcommand{\snP}{{\rm P}}
\newcommand{\snD}{{\rm D}}

\newcommand{\ud}{\mathrm{d}}

\def\bse{\begin{subequations}}
\def\ese{\end{subequations}}

\begin{document}
\title{A high flux source of cold strontium atoms}

\author{T.~Yang\inst{1}\thanks{\emph{Present address:}{NIM, Beijing, P.R.China}} \and K.~Pandey\inst{1} \and M.S.~Pramod\inst{1}  \and F.~Leroux \inst{1} \and C.C.~Kwong\inst{2} \and E.~Hajiyev\inst{1} \and Z.Y.~Chia\inst{1}\thanks{\emph{Present address:}{Earth Observatory of Singapore, NTU, Singapore}} \and B.~Fang\inst{1}\thanks{\emph{Present address:}{SYRTE, Paris, France}} \and D.~Wilkowski\inst{1}\inst{2}\inst{3}\thanks{\emph{Corresponding author:}{david.wilkowski@ntu.edu.sg}}%
}                     
\institute{Centre for Quantum Technologies, National University of Singapore , 117543 Singapore, Singapore. \and PAP, School of Physical and Mathematical Sciences, Nanyang Technological University, 637371 Singapore, Singapore. \and MajuLab, CNRS-University of Nice-NUS-NTU International Joint Research Unit UMI 3654, Singapore.}
\date{Received: date / Revised version: date}
%
\abstract{
We describe an experimental apparatus capable of achieving a high loading rate of strontium atoms in a magneto-optical trap operating in a high vacuum environment. A key innovation of this setup is a two dimensional magneto-optical trap deflector located after a Zeeman slower. We find a loading rate of \unit{6\times 10^9}{s^{-1}} whereas the lifetime of the magnetically trapped atoms in the $^3\snP_2$ state is \unit{54}{s}.}
%
\maketitle
\section{Introduction}
\label{sec_intro}

Cooling and trapping of alkaline-earth atoms, and in the particular case of strontium atoms, have been rapidly developed over the last two decades~\cite{PhysRevLett.82.1116,Katori,doi:10.1142/S0217984906011682,0256-307X-26-9-093202,PhysRevLett.103.200401,PhysRevLett.103.200402,PhysRevLett.107.243002,stellmer2013laser,1063-7818-45-2-166}. These two valence electrons atoms have a ground state with zero electronic spin and weakly allowed singlet-triplet transitions. They offer interesting alternatives to the more commonly used alkali atoms and open new fields of research for cold and ultracold gases. In the field of ultra high precision spectroscopy, tremendous progress has been made recently to reduce the frequency instabilities and systematic shifts of atomic clocks. Currently, the best results are obtained with strontium lattice clocks (see \cite{KatoriLattice,Clock87Sr} for first implementations and \cite{RevModPhys.83.331,LBY2015,POG2013} for reviews) where the accuracy and clock frequency difference were measured at the $10^{-18}$ level~\cite{bloom2014optical,ushijima2014cryogenic}. Those developments are of importance not only for time keeping, but also for precision tests in fundamental physics~\cite{Chou24092010,Rosenband28032008,akamatsu2014frequency}. Using different strontium isotopes, test of the equivalence principle and search for spin-gravity coupling effects have been reported~\cite{PhysRevLett.113.023005}. In studies related to quantum gases, production of a $^{84}\textrm{Sr}$ Bose-Einstein condensate (BEC) was achieved by the Innsbruck and the Rice groups~\cite{PhysRevLett.103.200401,PhysRevLett.103.200402}. These were rapidly followed by the production of quantum gases of different Sr isotopes and their mixtures~\cite{desalvo_Sr,PhysRevA.81.051601,PhysRevA.82.041602,PhysRevA.82.011608}. Bosonic strontium atoms are textbook examples of scalar BEC, whereas the $^{87}\textrm{Sr}$ Fermion has a large $I=9/2$ nuclear spin that allows the study of Kondo physics~\cite{PhysRevA.81.051603}, SU(N) magnetism \cite{zhang2014spectroscopic} and gauge field~\cite{PhysRevLett.110.125303,PhysRevA.90.023601}. The lack of electronic and nuclear spin in the ground state of the bosonic isotopes reduces the resonant photon scattering phenomena to its simplest component: Rayleigh scattering (\emph{i.e.}, no Raman scattering). In this context, it has been shown that the photon coherence is preserved during multiple photon scattering~\cite{PhysRevLett.88.203902}. Furthermore, it is destroyed only by the intensity saturation effect~\cite{PhysRevE.70.036602}.  Finally, the narrow intercombination line of $^{88}\textrm{Sr}$ was used to reveal cooperativity and (super)flash effects in forward scattering of an optically thick cold cloud~\cite{PhysRevA.84.011401,PhysRevLett.113.223601}.

In this article, we describe a new experimental apparatus that achieves a high loading rate of a strontium magneto-optical trap (MOT) in a high vacuum environment. Indeed, we measure a loading rate of \unit{6\times 10^9}{s^{-1}} which is, to the best of our knowledge, the highest rate reported for a strontium setup. Such a high rate considerably reduces the loading time of the MOT. This is helpful to decrease the duty cycle in optical lattice clocks, minimizing the Dick effect (see for example \cite{danet2014dick} and references therein), or to produce degenerate quantum gases with higher number of atoms~\cite{stellmer_2013}. High loading rates also lead to samples that are optically thicker, which are important to explore cooperative phenomena~\cite{PhysRevLett.113.223601}.

\begin{figure}[h]
\includegraphics[width=0.48\textwidth]{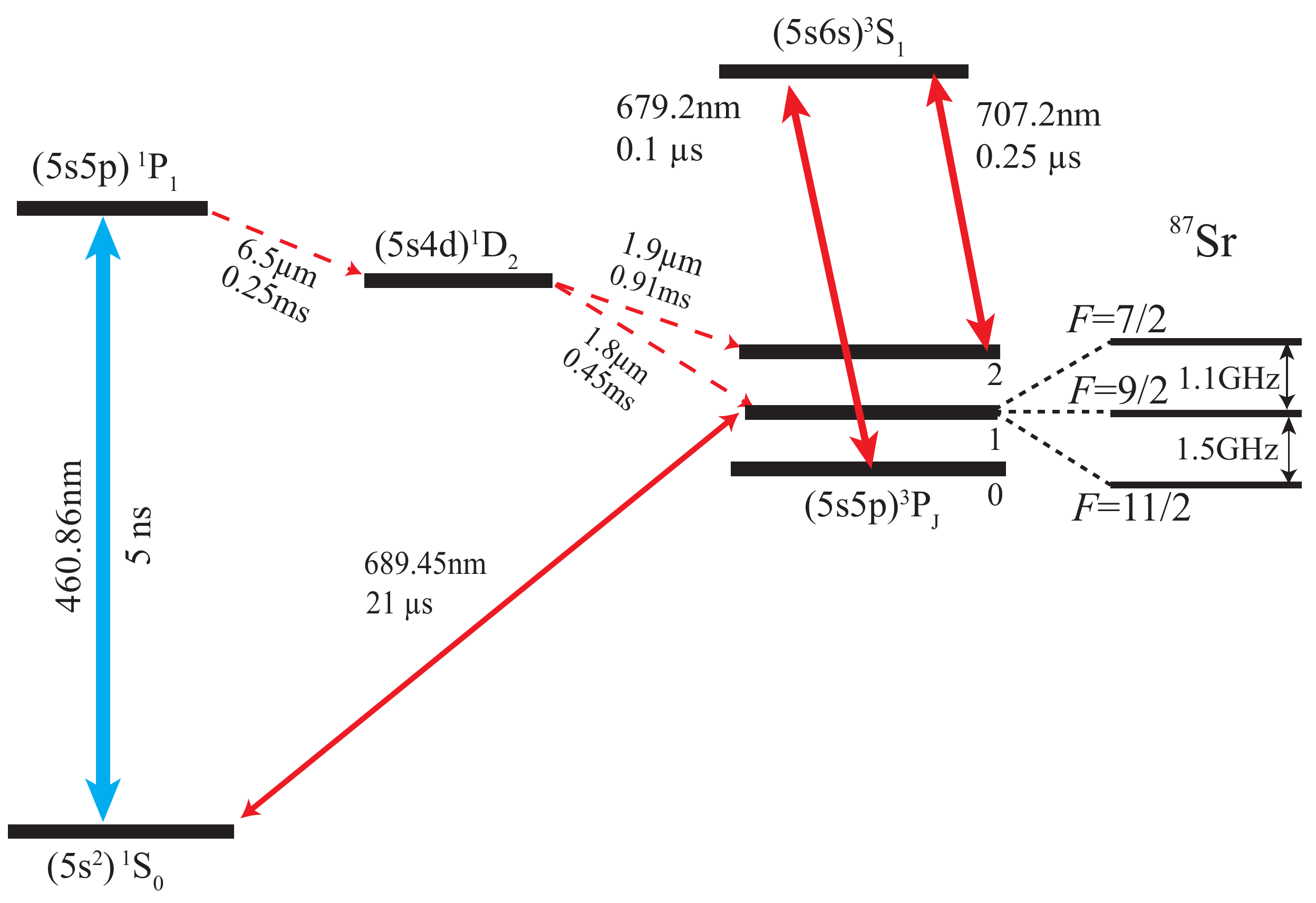}
\caption{Energy levels and transitions of interest of strontium. The excited states decay time and wavelength are indicated on the graph.}
\label{Srtransitionlevel}
\end{figure}

We describe the different components of our experimental apparatus in section~\ref{sec_exp_apparatus}. One key feature is a two dimensional MOT deflector located after a Zeeman slower (sections~\ref{subsec_zeeman_slower} and~\ref{subsec_2D_deflector}). It increases the brightness of the slow beam of atoms and deflects it by an angle of 30$^\circ$. The beam is redirected into a differential pumping tube before it enters into the science chamber (section~\ref{subsec_3D_blue_MOT}). This setup allows us to maintain a low residual background pressure at the position of the MOT, resulting in a \unit{54}{s} lifetime of the $^3\snP_2$ state magnetic quadrupole trap. The details of the laser systems addressing the various transitions (shown in Fig.~\ref{Srtransitionlevel}) are given in section~\ref{sec_laser_system}. In particular, the laser system, operating on the $^1\snS_0\rightarrow \,^1\snP_1$ dipole allowed transition at \unit{461}{nm}, is locked on a separate Sr atomic beam. We tune the frequency lock point by varying the magnetic field. This allows us to address the different strontium isotopes (section~\ref{sub_sec_blue_laser}). Finally, we modulate the phase of both the \unit{707}{nm} and \unit{679}{nm} repumping lasers. We generate frequency sidebands which address, at the same time, $^{88}\textrm{Sr}$ and the complete hyperfine structure of $^{87}\textrm{Sr}$. This aspect is discussed in section~\ref{sub_sec_repumper_laser}.

\section{Experimental apparatus}
\label{sec_exp_apparatus}

The relevant energy levels and transitions for laser cooling and trapping of strontium atoms are shown in Fig.~\ref{Srtransitionlevel}. From the singlet ground state $^1\snS_0$, there are two cooling transitions. The first one is the blue dipole allowed transition $^1\snS_0\rightarrow\,^1\snP_1$ at \unit{\lambda_b=461}{nm}. Its linewidth is \unit{\Gamma_b=2\pi\times 32}{MHz} and the saturation intensity is \unit{42.5}{mW cm^{-2}}. This transition is used to manipulate the effusive atomic beam (sections~\ref{subsec_2D_molasses}, \ref{subsec_zeeman_slower} and~\ref{subsec_2D_deflector}).  It is also used in the first stage of laser cooling in a magneto-optical trap (section~\ref{subsec_3D_blue_MOT}). The other cooling transition is the red intercombination line $^1\snS_0\rightarrow\,^3\snP_1$ at \unit{\lambda_r=689}{nm}. Its linewidth \unit{\Gamma_r=2\pi\times7.5}{kHz} is much narrower than the blue transition. Its saturation intensity \unit{3}{\micro W cm^{-2}} is also much weaker. This transition is used in the final cooling stage for both the bosonic $^{88}\textrm{Sr}$ and the fermionic $^{87}\textrm{Sr}$ isotopes (section~\ref{subsec_red_MOT_88}).

\begin{figure}[h]
\includegraphics[width=0.48\textwidth]{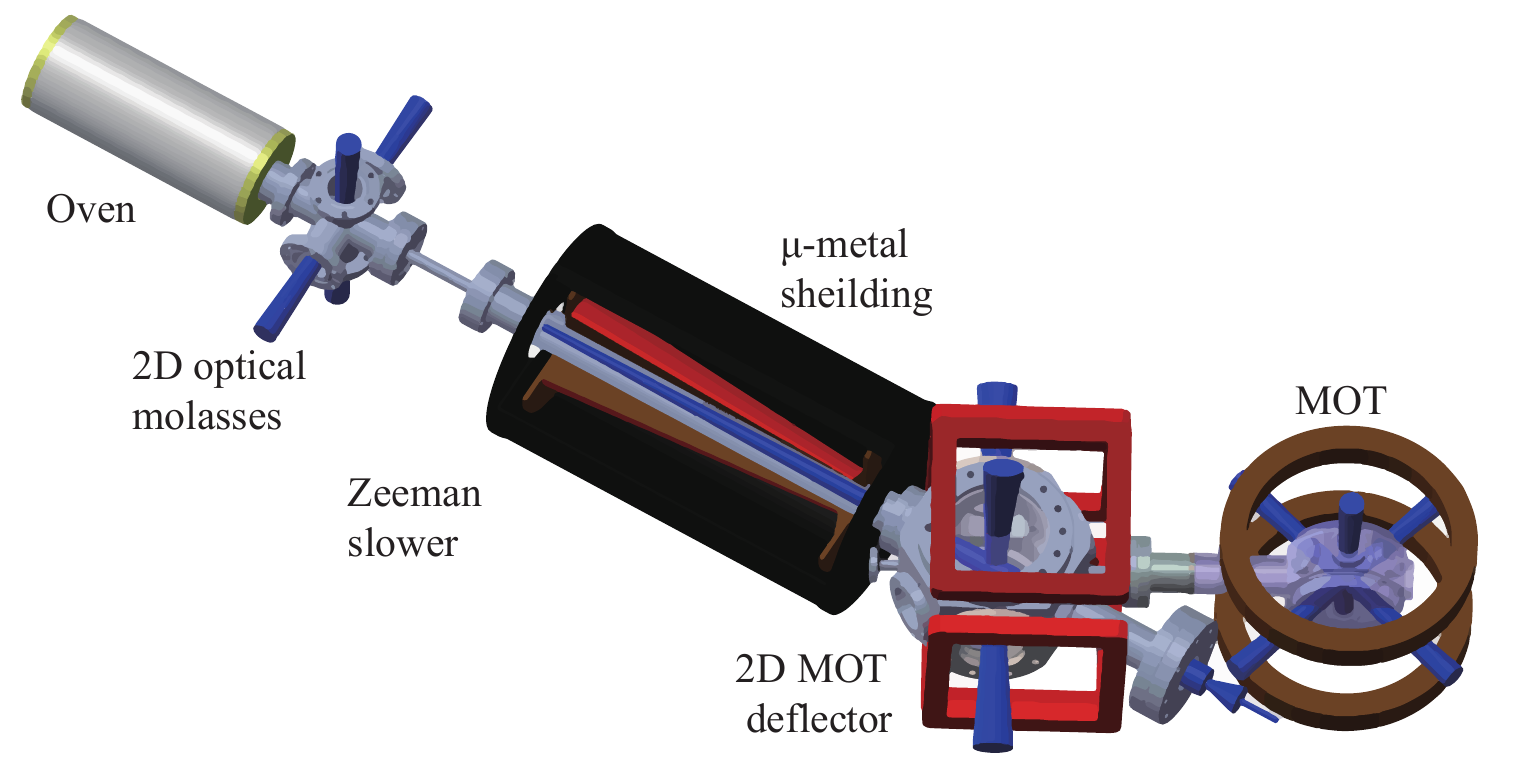}
\caption{Three dimensional schematic view of the vacuum apparatus.}
\label{SysAppara3D}
\end{figure}

A sketch of the experimental setup is shown in Fig.~\ref{SysAppara3D}. It consists of an oven and four stages of cooling and manipulation of the atomic beam, namely, a two dimensional optical molasses (2D OM), a Zeeman slower, a two dimensional magneto-optical trap (2D MOT), and finally a magneto-optical trap (MOT). The effusive atomic beam, extracted from the oven, is first transversely cooled by the 2D OM before entering the Zeeman slower, where its longitudinal velocity is reduced to \unit{~60}{m/s}. This value is within the velocity capture range of the MOT. Finally, the atoms are deflected and guided into the MOT chamber by a 2D MOT. All these preparatory stages, described in detail in the following sections, use the broad blue dipole allowed transition.

Excluding the strontium oven, the other elements of our apparatus are analyzed and optimized in terms of the atom loading rate $L_0$ into the MOT. $L_0$ is extracted from measurements of the fluorescence signal of the cold gas. For this purpose, we use a resonant \unit{10}{mW} blue probe beam with a waist of \unit{3.5}{cm} that is  larger than the cold cloud size. From the fluorescence signal, we deduce the number of cold atoms in the MOT. To avoid systematic errors in the atom counting, we calibrate the fluorescence signal with the absorption imaging, by a CCD camera, of a cold cloud containing a small number of atoms, using the same probe. The loading rate values are extracted from a fluorescence measurement, done \unit{5}{ms} after turning on the MOT beams. Since this time is much shorter than the characteristic MOT lifetime (section~\ref{subsec_3D_blue_MOT}), the number of atoms in the MOT increases linearly. Moreover, the maximum number of atoms in the MOT during the loading rate measurement is $3\times 10^7$ with a density that is lower than \unit{10^9}{cm^{-3}}. Under these conditions, we do not expect any bias on the atom counting due to multiple photon scattering~\cite{labeyrie2003slow}.

\subsection{Strontium oven}
\label{subsec_strontium_source}
The oven is made of a stainless-steel cylinder heated at a maximal temperature of \unit{550}{ ^\circ C}. This temperature is stabilized to ensure a constant flux of the outgoing effusive atomic beam. The beam is generated in a multichannel nozzle made of nine hundred Monel400 micro tubes clamped in a \unit{1 \times 1}{cm} square area. The length of each tube is \unit{1}{cm} with inner and outer diameters of \unit{200}{\mu m} and \unit{300}{\mu m} respectively. We keep the nozzle at a temperature of roughly \unit{50}{ ^\circ C} above the oven temperature to avoid clogging the tubes.

\begin{figure}[h]
\includegraphics[width=0.48\textwidth]{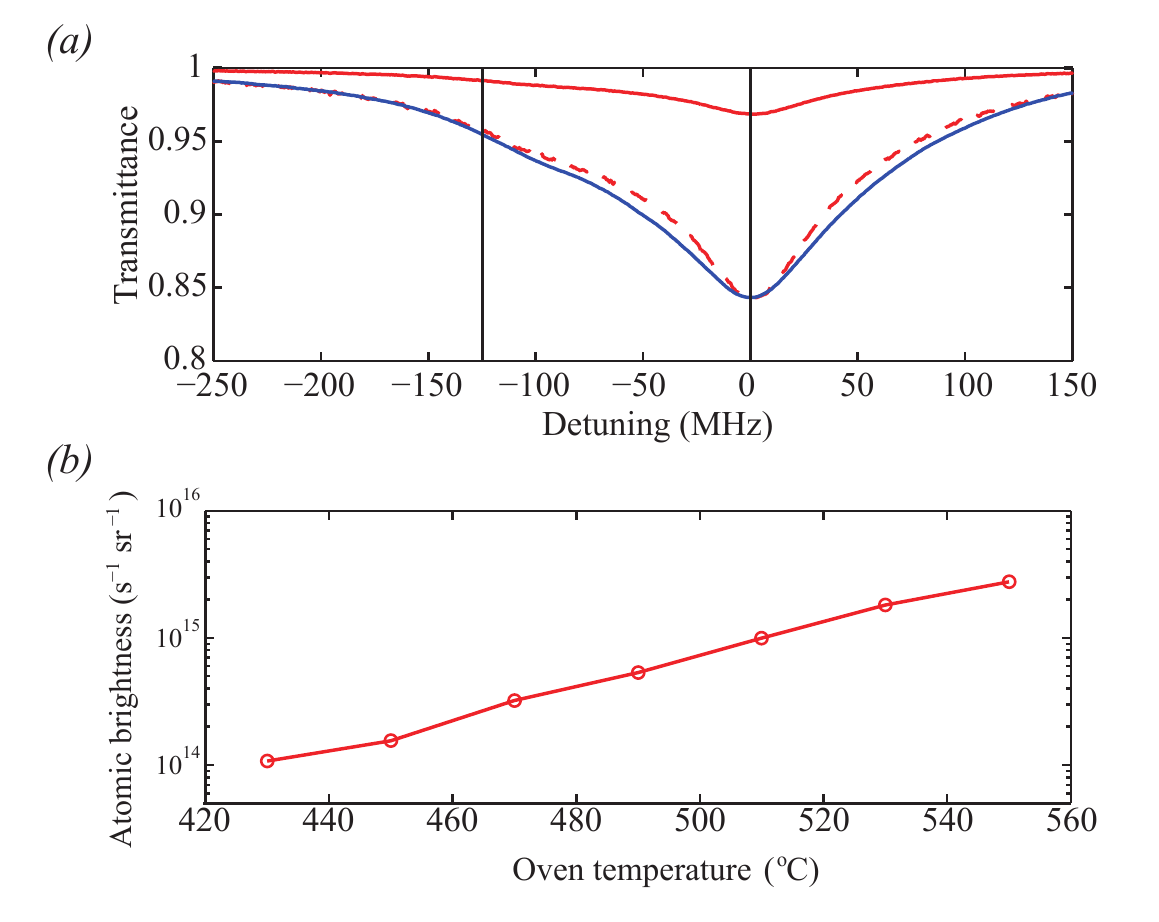}
\caption{(a) Transverse spectra of the atomic beams after the oven. The solid red (blue) curve corresponds to a temperature of \unit{T=490}{^\circ C} (\unit{T=550}{^\circ C}). To compare both profiles, we rescale the low temperature curve such that both spectra show the same amplitude. This rescaling operation is represented by the dashed red curve. The two vertical solid lines indicate the position of the $^{88}$Sr and $^{86}$Sr resonances at \unit{0}{MHz} and \unit{-125}{MHz} respectively. (b) Atomic beam brightness as a function of the oven temperature.}
\label{OvenSide}
\end{figure}

In Fig.~\ref{OvenSide}(a), we give two examples of the transverse transmission spectra of the atomic beam. These spectra are taken at \unit{10}{cm} after the oven nozzle. They reflect the transverse velocity distribution of the atomic beam. The velocity distributions have a triangular-like shape, a characteristic of the collisionless regime inside the nozzle micro tubes~\cite{pauly2000atom}. As a result, we observe that the brightness of the atomic beam increases exponentially in the temperature range of interest [Fig.~\ref{OvenSide}(b)].

\subsection{Two dimensional optical molasses}
\label{subsec_2D_molasses}
\begin{figure}[h]
\includegraphics[width=0.48\textwidth]{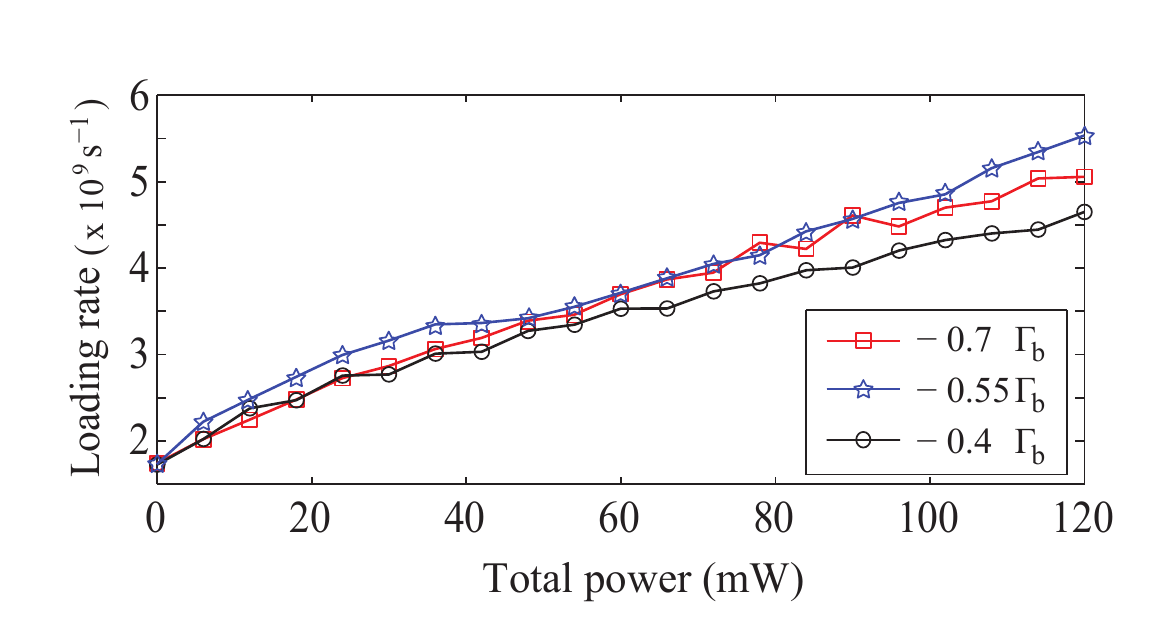}
\caption{MOT loading rate as a function of the 2D OM total optical power for different detunings indicated on the graph. Since the beams are retro-reflected, the total power is twice the incident power.}
\label{2DOM}
\end{figure}

The effusive atomic beam, produced in the oven as explained in the previous section, is transversely cooled by a two dimensional optical molasses (2D OM) located \unit{15}{cm} after the nozzle. We use two pairs of retroreflected beams having the same elliptical shape (beam waists: \unit{1}{cm} and \unit{0.5}{cm}). The major axis of the ellipse is oriented along the atomic beam. The performance of the 2D OM cooling is measured in terms of the atom loading rate $L_0$ into the MOT. This quantity is plotted in Fig.~\ref{2DOM} for different values of the frequency detuning and the total power of the cooling beams. The optimum performance is found for a detuning of \unit{-18}{MHz}$\,=-0.55\Gamma_b$. It corresponds to a three-fold improvement of the MOT loading rate. We note that further improvements may be expected at a higher laser power. Here, we are limited by the total available laser power (section~\ref{sub_sec_blue_laser}).

After the 2D OM, the transversely cooled atoms pass through a \unit{10}{cm} long tube with an inner diameter of \unit{1}{cm}. The main purpose of this tube is to block the unwanted atoms with high transverse velocities. By this means, we also create a differential pressure environment between the oven and the subsequent stages of the vacuum apparatus.

\subsection{Zeeman slower}
\label{subsec_zeeman_slower}
The longitudinal velocity of the atomic beam is reduced using a \unit{32}{cm} long Zeeman slower (see Fig.~\ref{SysAppara3D}). The inhomogeneous longitudinal magnetic field profile of the Zeeman slower is created by winding flat copper wires in a tapered way. At the optimum electrical current of \unit{6.6}{A}, the magnetic field of the Zeeman slower varies from \unit{500}{G} at the entrance to \unit{-100}{G} at the output. The optical beam of the Zeeman slower has a frequency detuning of $-13\Gamma_b$ and a power of \unit{\sim 110}{mW}. The laser beam enters into the apparatus through an optical grade \emph{z}-cut sapphire viewport. We permanently heat the viewport at a temperature of \unit{375}{^\circ C} to prevent the deposition of Sr atoms.

\begin{figure}[h]
\includegraphics[width=0.48\textwidth]{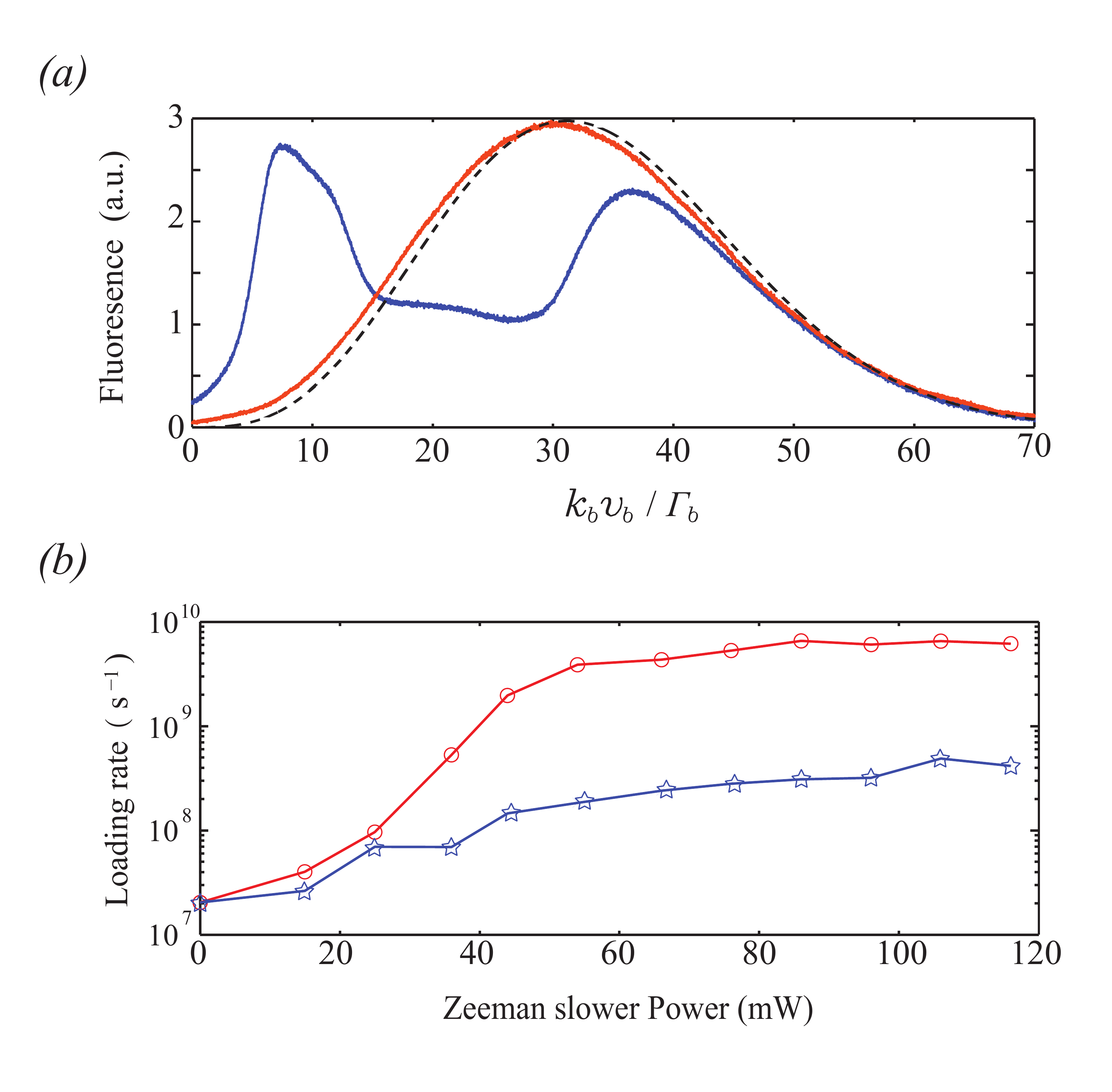}
\caption{(a) Atomic longitudinal velocity distribution measured after the Zeeman slower. We use a dimensionless velocity unit where \unit{\Gamma_b/k_b=15}{m/s}. The red solid line is the atomic velocity distribution without the Zeeman slower optical beam. The black dashed line is the theoretical distribution for an oven at $T=450^\circ$C. The blue line is the velocity distribution with the Zeeman slower operating optimally. (b) MOT loading rate as function of the Zeeman laser power when the magnetic field is turned on (red dots and solid lines) and when it is turned off (blue stars and solid lines). }
\label{Zeeman}
\end{figure}

We scan the frequency of a probe around the blue resonance to measure the longitudinal velocity distribution of the atomic beam  at the output of the Zeeman slower. The probe propagation axis is at an angle of $70^\circ$ with respect to the atomic beam, limiting the velocity resolution to $\sim \pm 3\Gamma_b/k_b$. Here, $\Gamma_b/k_b=15$~m/s, and $k_b=2\pi/\lambda_b$ is the wavevector of the probe. Some examples of the velocity distribution are given in Fig.~\ref{Zeeman}(a). When the Zeeman slower is turned off, the velocity distribution is in good agreement with the expected longitudinal distribution calculated for an oven at $T=450~^\circ$C. Once the Zeeman slower is turned on, the velocity of a large fraction of the atoms is brought down, contributing to a peak at $\sim 7\Gamma_b/k_b$ [see the blue curve in Fig.~\ref{Zeeman}(a)]. We note that the Zeeman slower is effective for velocities lower than $\sim 35\Gamma_b/k_b$. This value is in agreement with the Zeeman shift imposed by the magnetic field at the entrance.

In Fig.~\ref{Zeeman}(b), we show in detail the performance of the Zeeman slower by plotting the loading rate of the MOT as a function of the optical power of the Zeeman slower. Two situations are analyzed; when the Zeeman slower magnetic field is off (blue stars) and when an optimum magnetic field profile is applied (red dots). At low power, \emph{i.e.} \unit{<30}{mW}, the two situations are equivalent, meaning that the weak radiation pressure force is not enough to impose a deceleration of the atoms compatible with the variation of the magnetic field inside the Zeeman slower. The Zeeman slower is thus inoperative. Above a power of \unit{60}{mW}, the Zeeman slower reaches its optimum performance, showing a $20$-fold improvement of the MOT loading rate compared to the situation where the magnetic field is turned off. Overall, the Zeeman slower improves the loading rate in the MOT by a factor of $300$.

The magnetic field of the Zeeman slower is confined at its vicinity using f-metal sheets (see Fig.~\ref{SysAppara3D}). This magnetic field shielding box does not affect the performance of the Zeeman slower. Rather, it prevents a distortion of the magnetic field lines of the two dimensional atomic beam deflector located just after the Zeeman slower.

\subsection{Two dimensional magneto-optical trap and beam deflector}\label{2DMOTD}
\label{subsec_2D_deflector}

\begin{figure}[h]
\includegraphics[width=0.48\textwidth]{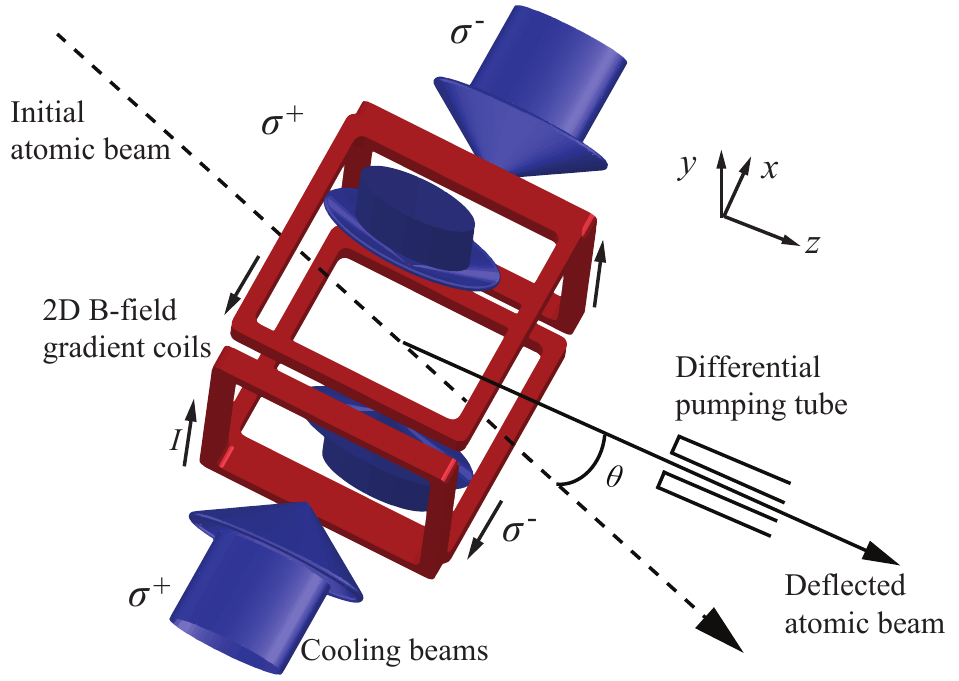}
\caption{Schematic view of the 2D MOT atomic beam deflector. The dashed line gives the direction of the initial atomic velocity. Two pairs of coils generate a 2D quadrupole field with a line of zero magnetic field along the $z$-axis at a $\theta=30^\circ$ angle with the incident atomic beam. The $z$-axis is also the outgoing direction from the 2D MOT. The two pairs of the retroflected MOT beams ($\sigma ^{+}$ - $\sigma ^{-}$ polarizations) are perpendicular to each other in the $x-y$ plane. After the 2D MOT, the deflected atomic beam passes through a differential pumping tube. }
\label{2DMOT}
\end{figure}

After the Zeeman slower, the atoms enter a two dimensional magneto-optical trap (2D MOT). This device aims to transversely cool, focus and guide the atomic beam along the zero magnetic field line located along its symmetry axis~\cite{berthoud,chen_Rb,Kai_2DMOT}. To deflect the atomic beam, the 2D MOT axis is purposely kept at an angle of 30$^\circ$ off the Zeeman slower axis , in the horizontal plane (see Fig.~\ref{SysAppara3D} and \ref{2DMOT}). At the output of the 2D MOT, the atomic beam passes through a differential pumping tube with an entrance radius of \unit{2.5}{mm} and an acceptance angle of \unit{10}{mrad}.

Under this original arrangement, our 2D MOT has at least four advantages. First, by deflecting the atomic beam, we prevent the intense Zeeman optical beam to cross, and thus to disturb, the MOT in the science chamber. Second, the uncooled atoms in the atomic beam, including the unwanted Sr isotope, do not enter the science chamber. Combined with the differential pumping tube, it allows us to isolate the principal part of our experimental setup, namely the science chamber, from the rest of the vacuum apparatus. Hence, we get an ultra high vacuum environment in the science chamber. This allows us to achieve a long lifetime of our cold sample for future evaporative cooling towards degenerate gases. Third, we gain an additional optical axis in the science chamber along the atomic beam.  Finally, once the MOT is loaded, the 2D MOT beams can be turned off so that the atomic flux, entering into the science chamber, drops to zero. Hence, this design ensures a high atomic flux, but does not require any extra mechanical blocker for the atomic beam.

A schematic view of the 2D MOT is given in Fig.~\ref{2DMOT}. The two dimensional quadrupole magnetic field is generated using four square shaped coils. The inner and outer sizes of the coil are \unit{15}{cm} and \unit{19}{cm} respectively. Each coil is formed by 22~turns $\times$ 20~layers of {\unit{1}{mm}} diameter copper wire. The distance between each parallel pair of coils is \unit{20}{cm}. This configuration gives a homogenous magnetic field gradient of $\partial B/\partial x = 5~$G/cm in the interacting zone for $1$~A of current. The cooling beams have an elliptical shape with an aspect ratio of $3\!:\!1$. The major axis is oriented along the atomic beam with a waist of \unit{3}{cm}. Using retro-reflected beams, the total available optical power is \unit{220}{mW}. After optimization, we found that $60\%$ of the total power has to be fed into the horizontal beam pair. Indeed, along this axis, larger radiation pressure forces are required to deflect the beam. The remaining power is sent into the vertical beam pair, where only the cooling and focusing of the atomic beam are required.

\begin{figure}[h]
\includegraphics[width=0.48\textwidth]{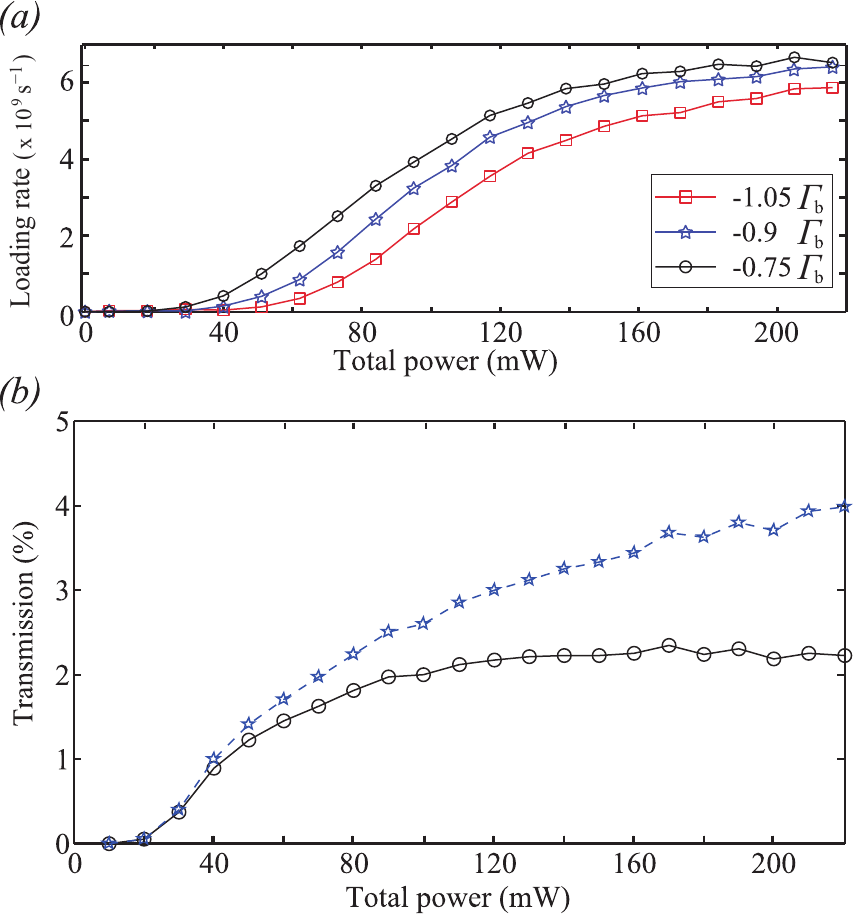}
\caption{(a) Loading rate of the MOT as a function of the total optical power of the 2D MOT for different frequency detuning indicated on the graph. Since the beams are retro-reflected, the total power is twice the incident power. The 2D MOT magnetic field gradient is \unit{7.5}{G/cm}. (b) Numerical simulation of the 2D MOT transmission percentage for a frequency detuning of $-0.9\Gamma_b$. The total number of atoms used in the simulation is $10^5$. The black open circles and the solid lines give the results considering optical pumping to the long lived state $^1\snD_2$. The case without optical pumping corresponds to the blue open stars and the dashed lines.}
\label{2DMOTPer}
\end{figure}

As we did for the 2D OM and the Zeeman slower, we analyze the performances of our 2D MOT beam deflector in terms of $L_0$, the loading rate in the MOT. In Fig.~\ref{2DMOTPer}(a), we show $L_0$ as a function of the total 2D MOT optical power for different detuning. The magnetic field gradient is set to the optimal value of \unit{7.5}{G/cm}. As a general feature, we observe a saturation of the loading rate above an optical power of \unit{~\sim 140}{mW}. Surprisingly, it occurs at a modest saturation intensity parameter of $0.16$. This value is calculated for a frequency detuning of $-0.9\Gamma_b$. We attribute this saturation behavior of the loading rate to a loss mechanism induced by spontaneous decay from the $^1\snP_1$ state to the underlying long-lived $^1\snD_2$ state. This optical pumping mechanism is known to reduce the blue MOT lifetime. In principle, this can be circumvented using repumping lasers, as it is successfully done for the MOT in the science chamber (see section~\ref{subsec_3D_blue_MOT} for more details). Unfortunately, it is inefficient here since the dwell time of the atoms in the 2D MOT is shorter than the time for repumping.

We perform numerical simulations to confirm the important role of the $^1\snP_1\rightarrow\,^1\snD_2$ spontaneous decay as a loss mechanism in the 2D MOT. Starting from realistic velocity and spatial probability distributions at the input of the 2D MOT, we compute the classical trajectory of an atom subjected to the following mean radiation pressure force: $\vec{F} = \left({F_{x+}} + {F_{x-}}\right)\hat{x}+\left({F_{y+}} + {F_{y-}}\right)\hat{y}$. The component of the force along $Ox$ axis is
\begin{equation}
{F_{x\pm} } =  \pm \frac{\hbar k\Gamma_b}{2}\frac{{I_x(y,z)/I_s}}{{1 + {{(2(\delta  \mp kv_x \mp \mu b'x/\hbar)/\Gamma_b )}^2}}},
 \label{AtomForc}
\end{equation}
where $b'$ is the magnetic field gradient and $I_x(y,z)$ the intensity of the corresponding beam. One has a similar expression for the components of the force along the $Oy$ axis, $F_{y\pm}$. We stop the simulation once the atom reaches entrance plane of the differential pumping tube. A successful transmission count fulfills two criteria. First, the atom must be located at the tube entrance within an acceptance angle, $\sqrt{v_x^2+v_y^2}/v_z$, smaller than \unit{10}{mrad}. Second, the atom should not have spontaneously decayed to the $^1\snD_2$ state. For this purpose we evaluate the probability of a decay using the transition rate $\overline{\pi}_P\Gamma_D$, where $\overline{\pi}_P$ is the averaged population of the excited $^1\snP_1$ state and \unit{\Gamma_D=4\times 10^3}{s^{-1}} is the $^1\snP_1\rightarrow^1\snD_2$ decay rate. Considering the transmission of the atoms through the 2D MOT with decay to the $^1\snD_2$ state included, the numerical simulation is in qualitative agreement with the experiment; it shows a saturation plateau for an optical power above \unit{~\sim 140}{mW} [see the black open circles and the solid lines in Fig.~\ref{2DMOTPer}(b)]. If the decay to the $^1\snD_2$ state is removed from the simulation, we observe a qualitatively different behavior, namely a larger transmission with a further increase at high optical power [see the blue open stars and the dashed lines in Fig.~\ref{2DMOTPer}(b)].

\subsection{Blue magneto-optical trap}
\label{subsec_3D_blue_MOT}

The atomic beam is transversely cooled, focused and deflected by the 2D MOT as described in section~\ref{subsec_2D_deflector}. After this stage, the atomic beam goes into the science chamber with a typical longitudinal mean velocity of $4\Gamma_b/k_b\sim 60$~m/s. This velocity is compatible with the velocity capture range of our MOT operating on the blue dipole allowed transition. It is, however, too high to be directly captured by the red MOT operating on the intercombination line (the typical capture velocity of the red MOT is \unit{1}{m/s}). Thus, the atoms are initially cooled and trapped in the blue MOT before being transferred into the red MOT for further gain in the phase-space density. The operation of the red MOT is described in the section~\ref{subsec_red_MOT_88}.

\begin{figure}[h]
\includegraphics[width=0.48\textwidth]{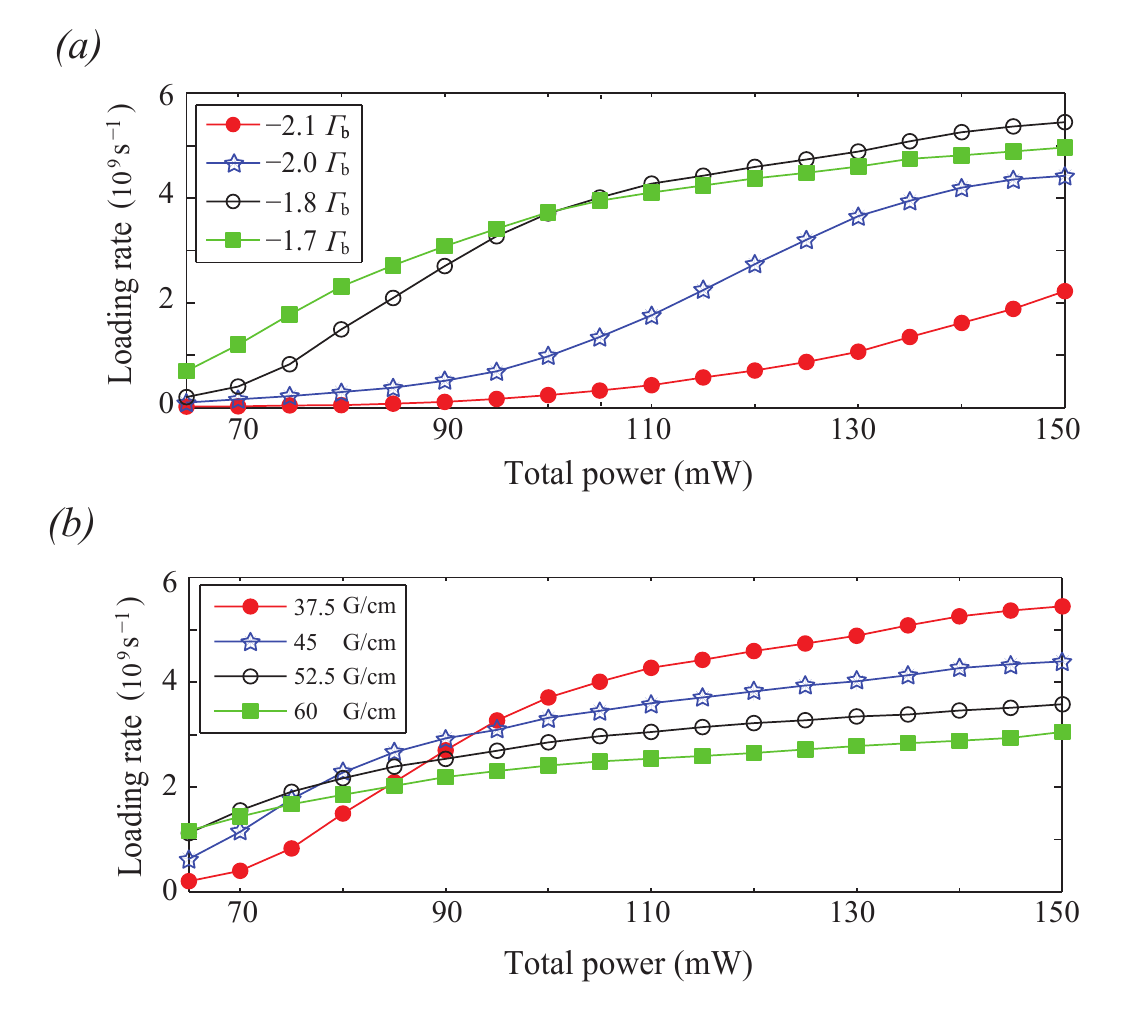}
\caption{The atomic loading rate in the MOT as a function of MOT beam total power. Since the beams are retro-reflected, the total power is twice the incident power. (a) Each curve corresponds to a different frequency detuning indicated on the graph. The magnetic field gradient is \unit{37.5}{G/cm}  (b) Each curve corresponds to a different MOT magnetic field gradient indicated on the graph. The frequency detuning of the MOT beams is $-1.8\Gamma_b$.}
\label{3DBlueMOT}
\end{figure}

Our blue MOT setup is made of three pairs of retroflected beams with a waist of \unit{1.1}{cm}, and a maximum total intensity of \unit{76}{mW cm^{-2}}. The typical performances of the blue MOT, in terms of the loading rate, are given in Fig.~\ref{3DBlueMOT}. We find a maximum atom loading rate of about \unit{6\times 10^9}{s^{-1}} at a magnetic field gradient of \unit{37.5}{G/cm} and a frequency detuning of \unit{-58}{MHz}.

\begin{figure}[h]
\includegraphics[width=0.48\textwidth]{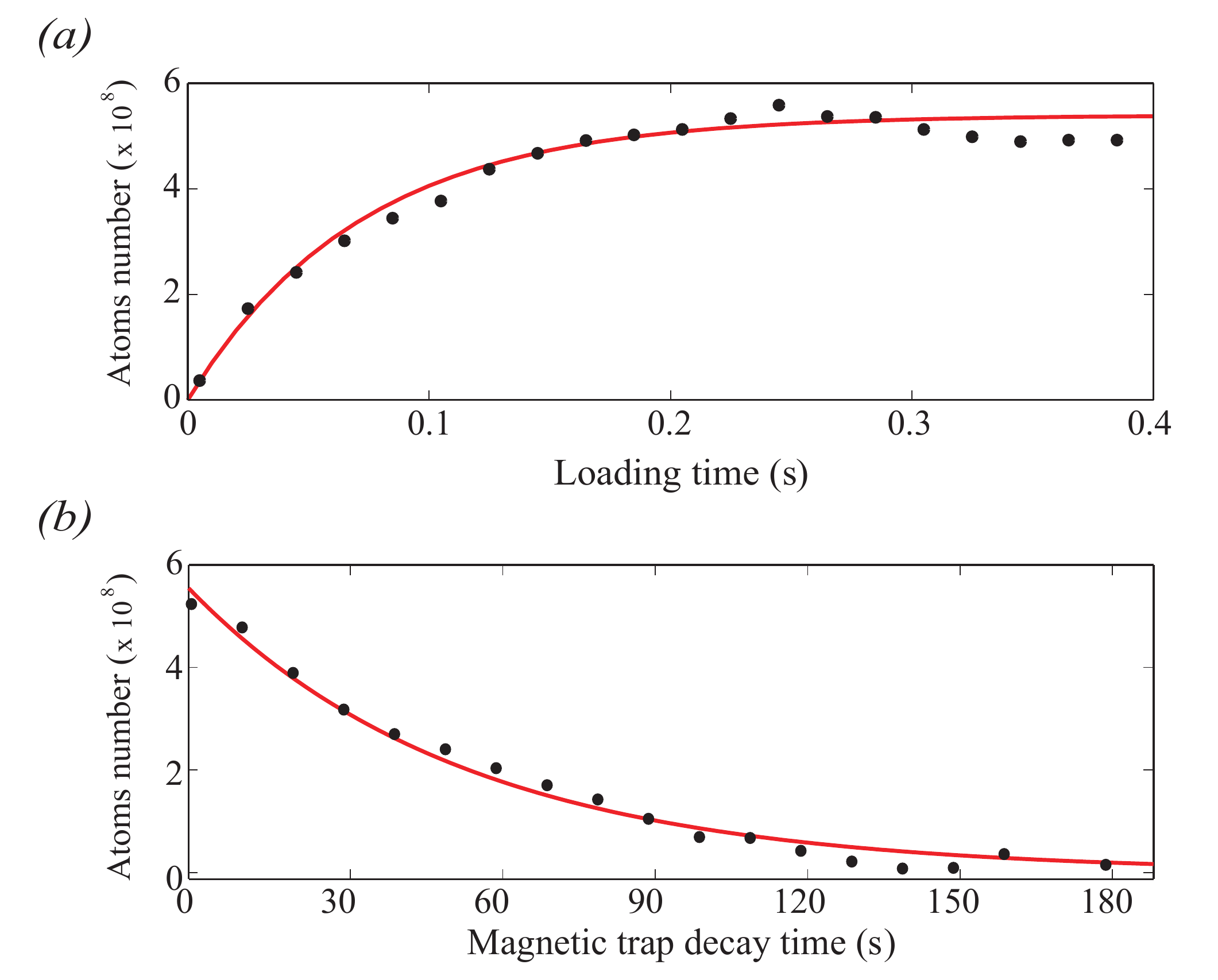}
\caption{(a) Temporal evolution of the number of $^{88}$Sr atoms in the blue MOT. The MOT beams are turned on at time $t=0$. (b) Temporal evolution of the atoms number magnetically trapped in the $^3\snP_2$ state after \unit{1}{s} of loading. The 2D MOT beams are turned off at \unit{t=-100}{ms}. For both graphs, the dots are the experimental values and the solid red lines are exponential fits of the data. For both graphs, the magnetic field gradient is set at \unit{45}{G/cm}.}
\label{lifetime}
\end{figure}

The lifetime of the cold atomic cloud depends strongly on the presence or absence of repumping lasers. Indeed, the $^1\snS_0\rightarrow^1\snP_1$ cooling transition is not perfectly closed because of the presence of an underlying $^1\snD_2$ state. As a result, atoms are lost by spontaneous decay into the long-lived triplet state $^3\snP_2$, with a typical MOT loss rate of \unit{50}{s^{-1}}. This value strongly depends on the $^1\snP_1$ state population, and therefore on the MOT beams intensity and frequency detuning. With repumping lasers (a detailed description is given in the section~\ref{sub_sec_repumper_laser}), this loss rate is reduced. An example of the temporal evolution of the atoms number in the MOT, after turning on the MOT beams, is shown in Fig.~\ref{lifetime}(a). Fitting the data with an exponential function, we find a loss rate of \unit{14}{s^{-1}}, limiting the atoms number in the MOT, at the stationary regime, to $N\simeq 5\times 10^8$. Such a high value of the loss rate is rather surprising. We believe it could be explained by unwanted escape channels in the MOT or incomplete repumping of the atoms from the metastable states. We shall exclude losses due to light assisted collisions since the two-body collisional loss coefficient reported in the literature is \unit{\beta\simeq2\times 10^{-10}}{cm^3/s}~\cite{dinneen1999cold,caires2004intensity}. With a peak spatial density of $n\,$\unit{\simeq 3\times 10^{9}}{cm^{-3}}, this leads to a small light assisted collision loss rate, estimated to be $\beta n\,$\unit{\simeq 0.6}{s^{-1}}. We exclude the background collision losses as well. This is inferred from the measurement of the atomic cloud lifetime in the $^3\snP_2$ state trapped in the magnetic quadrupole field. This experiment is done by first loading the MOT without repumping lasers for \unit{1}{s}. After this loading sequence, we hold the atoms for a variable time before turning on the repumping laser. The population in the magnetic trap is then transferred to the ground state and measured in the MOT. We find an atoms loss rate of \unit{18\times 10^{-3}}{s^{-1}} in the magnetic trap [Fig.~\ref{lifetime}(b)]. This loss rate is in agreement with loss mechanisms induced by the blackbody radiation at room temperature through the $^3\snP_2\,\rightarrow\,^3\snD_{1,2,3}$ transitions~\cite{PhysRevLett.92.153004}. This gives an upper limit to our background collision loss rate and confirms the good vacuum environment in the science chamber.

 Finally, we note that higher atoms number can be obtained using an atom shelving technique by switching off the repumping lasers. In this technique, the atoms are loaded in the long lived $^3\snP_2$ magnetic trap rather than in the blue MOT.~\cite{PhysRevLett.92.153004,stellmer2014reservoir,PDF2005}.

\subsection{Red magneto-optical trap}
\label{subsec_red_MOT_88}

The final laser cooling sequence is performed in a red MOT operating on the $^1\snS_0\rightarrow\,^3\snP_1$ intercombination line at \unit{689}{nm}. We use similar techniques to those reported in the early studies by the JILA and the Tokyo groups~\cite{PhysRevLett.82.1116,vogel1999narrow}. In preparation for the transfer into the red MOT, the temperature of the cold gas in the blue MOT is reduced by decreasing the power of the blue MOT cooling beam down to {\unit{5}{mW} in \unit{50}{ms}. The laser power is then kept at this value for \unit{100}{ms} (see the time sequence in Fig.~\ref{3DRedMOT88}). The final temperature in the blue MOT is about \unit{2}{mK}, corresponding to an \emph{rms} velocity of \unit{\bar{\it{v}}\approx 0.5}{m/s}~\cite{chaneliere_2005}. In the red MOT, the minimal stopping distance is $\bar{v}^2/(2|a|)\approx 1$~mm, where $|a|=\hbar k_r/(2\Gamma_r)\approx 155$~ms$^{-2}$ is the acceleration induced by the maximal radiation pressure force. This simple calculation shows that, first, the stopping distance can be shorter than the typical laser beam size (\unit{1.4}{cm} in our case). Second, it is crucial to maintain a large radiation pressure force across the initial velocity distribution of the cloud. For this purpose, the laser frequency linewidth is artificially broadened to \unit{2}{MHz}$\approx 3k\bar{v}$ by frequency modulation~\cite{chaneliere_2007}. The total power of the red MOT beams is \unit{30}{mW} and the modulation rate is \unit{25}{kHz}. Thus, the intensity per red MOT beam and on each comb is \unit{120}{\mu Wcm^{-2}}, which is 40 times the saturation intensity. This configuration ensures a large radiation pressure force during the transfer into the red MOT. Finally, the artificial broadening is turned off to optimize the temperature and/or the spatial density. In the following section, we discuss in further details, the operation of the red MOT for the $^{88}$Sr and $^{87}$Sr isotopes. The difference between these two cases comes from the presence of a Zeeman state manifold and a hyperfine structure in the excited state for the fermionic isotope, introducing more complexity in the cooling scheme.

\begin{figure}[h]
\includegraphics[width=0.48\textwidth]{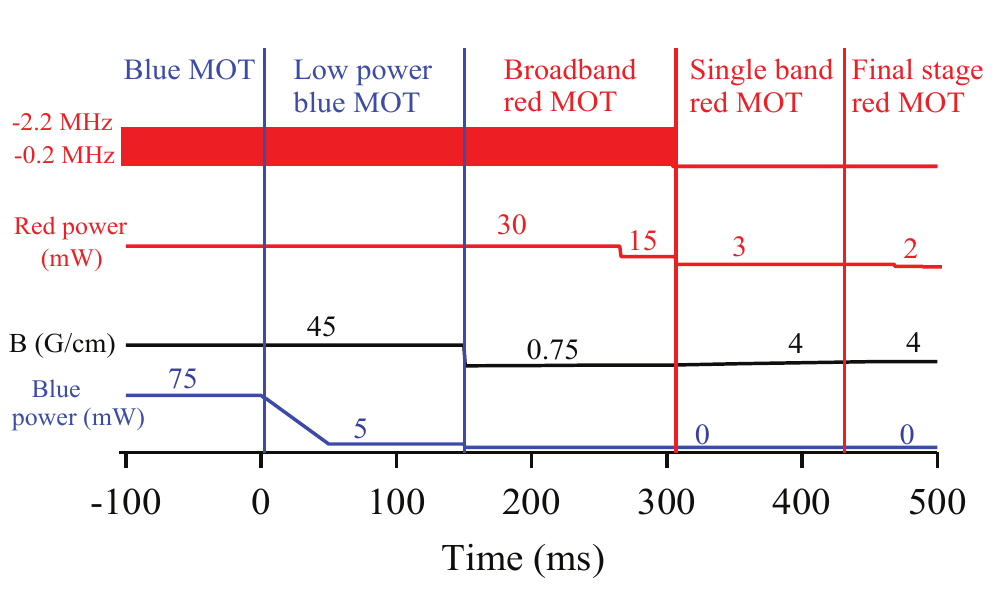}
\caption{Time sequence for the loading and cooling of $^{88}$Sr in the red MOT.}
\label{3DRedMOT88}
\end{figure}


\subsubsection{The $^{88}$Sr boson}
\label{subsub_sec_bosonic case}

The detailed time sequence of the red MOT for the $^{88}$Sr bosonic case is shown in Fig.~\ref{3DRedMOT88}. At the final stage of the broadband cooling, we typically transfer 50 percent of atoms into the red MOT. Then, the red MOT laser is switched to a single frequency with a detuning of \unit{-200}{kHz} $\approx -66\Gamma_r$. The MOT magnetic field gradient is ramped up from \unit{0.75}{G/cm} to \unit{4}{G/cm} in \unit{100}{ms} to compress the cold cloud. At the final stage of the cooling sequence, we find 3.5$\times 10^{8}$ atoms in a volume of \unit{0.15}{mm^3}, at the temperature of $T\,$\unit{=3}{\micro K}. The spatial density is $n\,$\unit{=2.5\times 10^{12}}{cm^{-3}}, leading to a phase space density of $3\times 10^{-3}$. Finally, we find a density dependance of the MOT temperature with a slope of $\ud T/\ud n\,$\unit{\simeq 6\times 10^{-12}}{cm^3 \micro K}. This shows that extra heating due to collective effects, such as multiple photon scattering, are still in play, even if the narrow transition has a linewidth in the same order of magnitude as the recoil frequency $\omega_r$, namely $\Gamma_r=0.6\omega_r$.

\subsubsection{The $^{87}$Sr fermion}
\label{subsub_sec_fermionic_case}

In contrast to the bosonic isotope, the $^{87}$Sr fermion has a nuclear spin $I=9/2$ which induces Zeeman sublevel manifolds and hyperfine splitting of the excited state. As a consequence, the MOT is less efficient and technically more challenging. We use the transition $F_g=9/2\rightarrow F_e=11/2$ to provide the necessary magneto-optical trapping force. However, because of the narrow resonance linewidth together with the presence of the MOT magnetic field, optical pumping among the Zeeman substates brings the atoms out of resonance. As initially proposed in Ref.~\cite{Katori}, we ensure proper remixing of the Zeeman substate by adding a stirring laser tuned on the transition $F_g=9/2\rightarrow F_e=9/2$. The stirring laser follows the same optical path as the MOT beams, thus it also participates in laser cooling.

Using acousto-optic modulators (AOMs), we control independently  the detuning and the intensity of the stirring and MOT lasers (see section~\ref{sub_sec_red_laser} for more details). The loading and cooling sequence of the $^{87}$Sr red MOT  is sketched in Fig.~\ref{RedMOT87}. After a blue MOT phase identical to that for the bosonic isotope, we first decelerate the atoms in an optical molasses for \unit{10}{ms}. Both stirring and MOT lasers are frequency broadened to \unit{2}{MHz} and brought to their maximum power, \emph{i.e.} \unit{25}{mW} and \unit{12}{mW} respectively. After the optical molasses stage, we apply a magnetic field gradient of \unit{2}{G/cm} for \unit{40}{ms} to compress the atomic cloud. Then, we reduce the laser powers to half of their initial values in \unit{30}{ms} in order to decrease the cloud temperature before switching to the single frequency operation. During this final cooling stage, the frequency detunings of the stirring and MOT lasers are \unit{-350}{kHz} and \unit{-220}{kHz} respectively, and their powers are progressively reduced to \unit{3}{mW} and \unit{2}{mW} respectively.  We get 3$\times$10$^7$~atoms at temperature of \unit{6}{\micro K} in a volume of \unit{\sim 1}{mm^3}. The reduction in the atoms number compared to the $^{88}$Sr case is consistent with the difference in the natural isotopic abundance. However, the larger cloud volume and the higher temperature indicate that cooling and trapping are less efficient for the fermionic isotope.

\begin{figure}[h]
\includegraphics[width=0.48\textwidth]{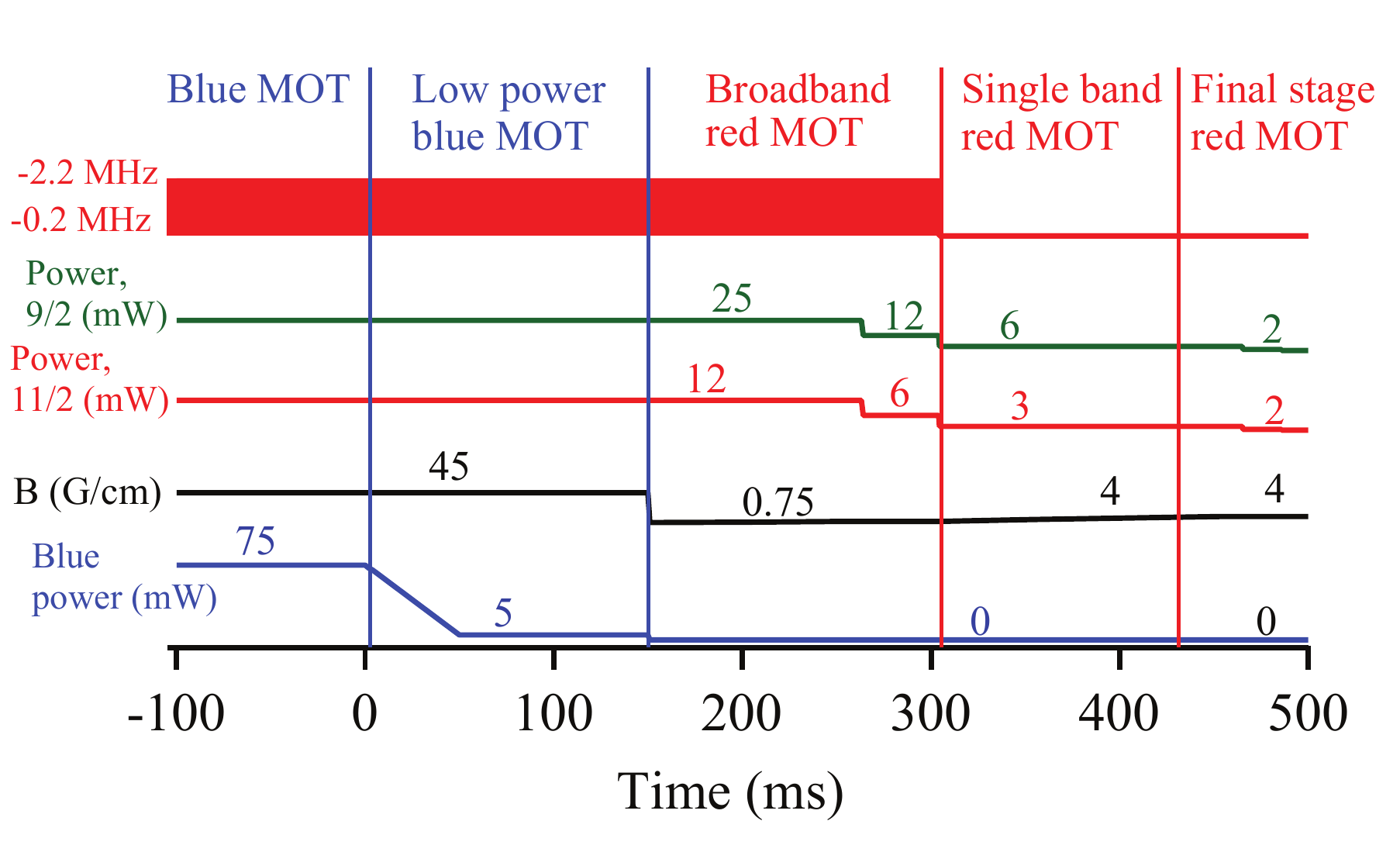}
\caption{Time sequence for the loading and cooling of $^{87}$Sr in the red MOT.}
\label{RedMOT87}
\end{figure}


\section{Laser systems}
\label{sec_laser_system}

This section is devoted to the description of the laser systems. The \unit{461}{nm} laser, addressing the $^1\snS_0\rightarrow\,^1\snP_1$ transition, is composed of a master laser at \unit{922}{nm}, a series of low and high power infrared (IR) optical amplifiers and two second harmonic generation (SHG) resonant cavities. The blue master laser is frequency locked on a $^{88}$Sr atomic beam. The lock point can be tuned via Zeeman effect using a pair of Helmholtz coils. This design allows us not only to change the global detuning of the laser, but also to address the different strontium isotopes (section~\ref{sub_sec_blue_laser}). The high frequency stability required for the $^1\snS_0\rightarrow\,^3\snP_1$ intercombination line at \unit{689}{nm} prevents us to implement a similar Zeeman based frequency tuning. Here, the different transitions of interest for $^{88}\textrm{Sr}$ and $^{87}\textrm{Sr}$ are addressed using AOMs and a \unit{1.2}{GHz} electro-optic modulator (EOM, see section~\ref{sub_sec_red_laser}). The red master laser is frequency locked on a high-finesse cavity. Its long term drift is compensated using a saturated fluorescence spectroscopy setup on an atomic beam. Finally, two lasers at \unit{707}{nm} and \unit{679}{nm} are tuned on the $^3\snP_2\rightarrow\,^3\snS_1$ and $^3\snP_0\rightarrow\,^3\snS_1$ transitions respectively for metastable state repumping during the blue MOT phase. With a single electro-optic modulator operating at high modulation index, we are able to address all the hyperfine sublevels of the $^{87}\textrm{Sr}$ $^3\snP_2$ state (section~\ref{sub_sec_red_laser}).

\subsection{Blue laser}
\label{sub_sec_blue_laser}
The optical setup of the blue laser system is shown in Fig.~\ref{BlueLaser}. Five output ports address the different parts of the experimental apparatus. AOMs and mechanical shutters are used to turn on and turn off the lasers beams. The primary laser is a commercial solid-state laser (Toptica TA-SHG Pro) which provides \unit{400}{mW} output at \unit{461}{nm}. It is composed of a \unit{922}{nm} external cavity diode laser (ECDL) used as a master laser, a \unit{1.5}{W} IR tapered amplifier (TA), and a SHG resonant cavity. A \unit{3}{mW} IR output port of the primary laser is used to inject a \unit{30}{mW} slave laser, after a single pass through two AOMs respectively at \unit{-45}{MHz} and \unit{-90}{MHz}. This laser beam injects a \unit{1}{W} TA after a \unit{60}{dB} isolator. Then, the beam is frequency doubled using a home-made bow type ring cavity containing a \unit{20}{mm} long periodically poled potassium titanyl phosphate (PPKTP) crystal. We get \unit{200}{mW} power at \unit{461}{nm}. This beam, used for the Zeeman slower, is red detuned by \unit{-2\times 135=-270}{MHz} with respect to the primary blue laser. Overall, the Zeeman slower beam is detuned by\unit{-408}{MHz} with respect to the frequency lock point.

The master laser is frequency locked using transmission spectroscopy at \unit{461}{nm} on an atomic beam extracted from an oven similar to the one described in section~\ref{subsec_strontium_source}. The error signal is obtained using a double pass AOM (AOM1 in Fig.~\ref{BlueLaser}), which is frequency modulated at a rate of \unit{10}{kHz}, with a modulation amplitude of {15}{MHz}. The large frequency scan generates an unwanted intensity modulation of the spectroscopy beam which is removed by a feedback loop on the laser intensity. The transmission signal of the laser beam through the atomic beam is collected on a photodetector before being demodulated and sent to a PID controller. The feedback loop of the lock is closed sending the error signal on the PZT of the master laser. We note that the atomic beam has an optical thickness of $0.7$. The width of the transition is \unit{100}{MHz}, which is Doppler broadened but remains reasonably narrow to provide an accurate frequency stabilization. The laser beam  is perpendicular to the atomic beam to remove any bias in the frequency due to Doppler shift. Finally, a pair of Helmholtz coils is placed across the atomic beam and aligned along the laser beam propagation axis [see Fig.~\ref{Isotopes}(a)]. Using a $\sigma^-$ polarized beam, the resonance frequency of the $^1\snS_0,m=0\rightarrow\,^1\snP_1,m=-1$ transition is shifted by Zeeman effect (the Land\'{e} factor of the excited state is $1$). We calibrate this shift as a function of the electric current passing through the coils and find the calibration coefficient to be \unit{36}{MHz/A}. This scheme allows us to scan the lock point of the laser across the strontium isotopes as it is illustrated in Fig.~\ref{Isotopes}(b).

\begin{figure}[h]
\includegraphics[width=0.48\textwidth]{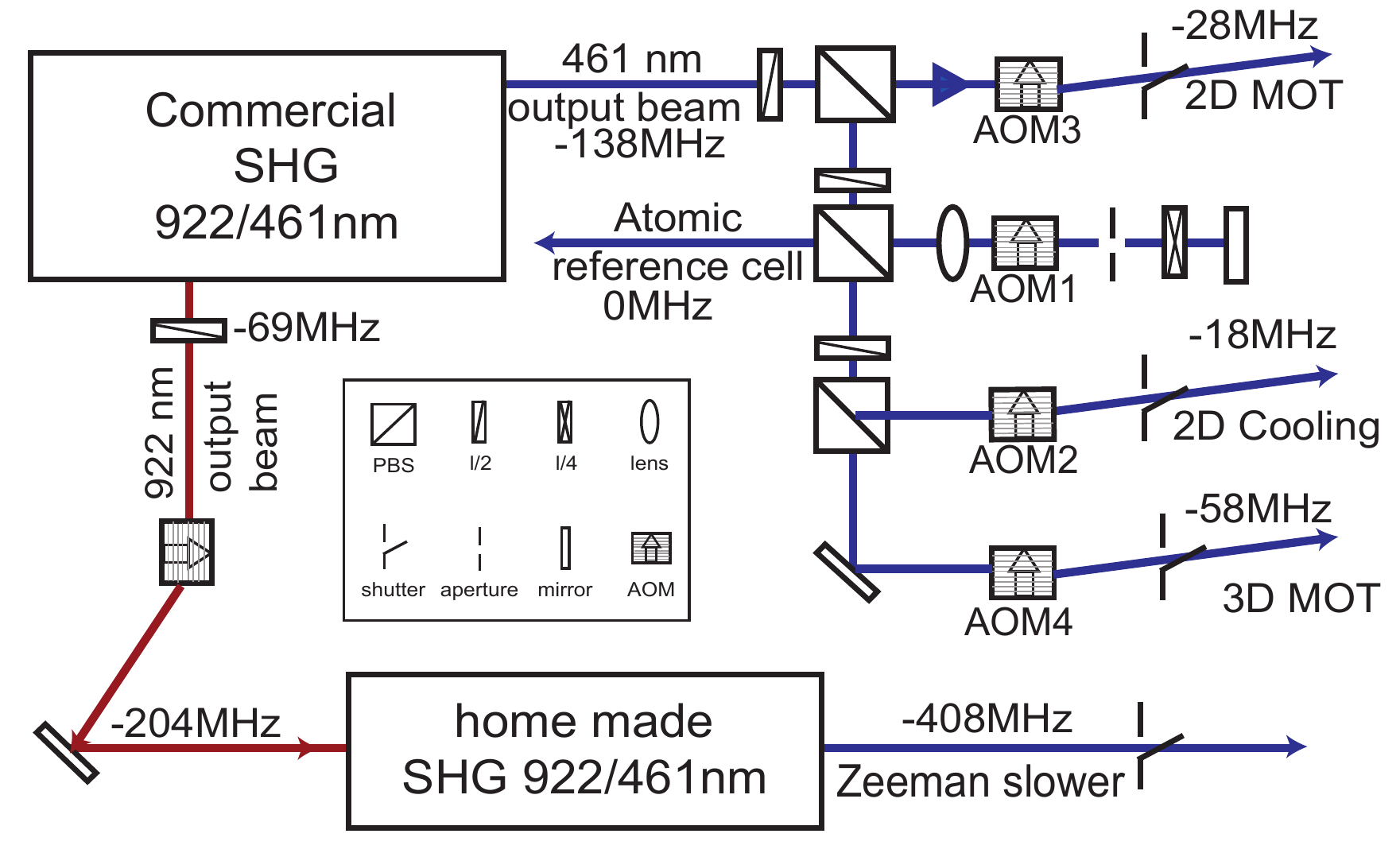}
\caption{Optical setup of the \unit{461}{nm} laser system. The frequencies of the different beams, given on drawing, are relative to the lock point.}
\label{BlueLaser}
\end{figure}


\begin{figure}[h]
\includegraphics[width=0.48\textwidth]{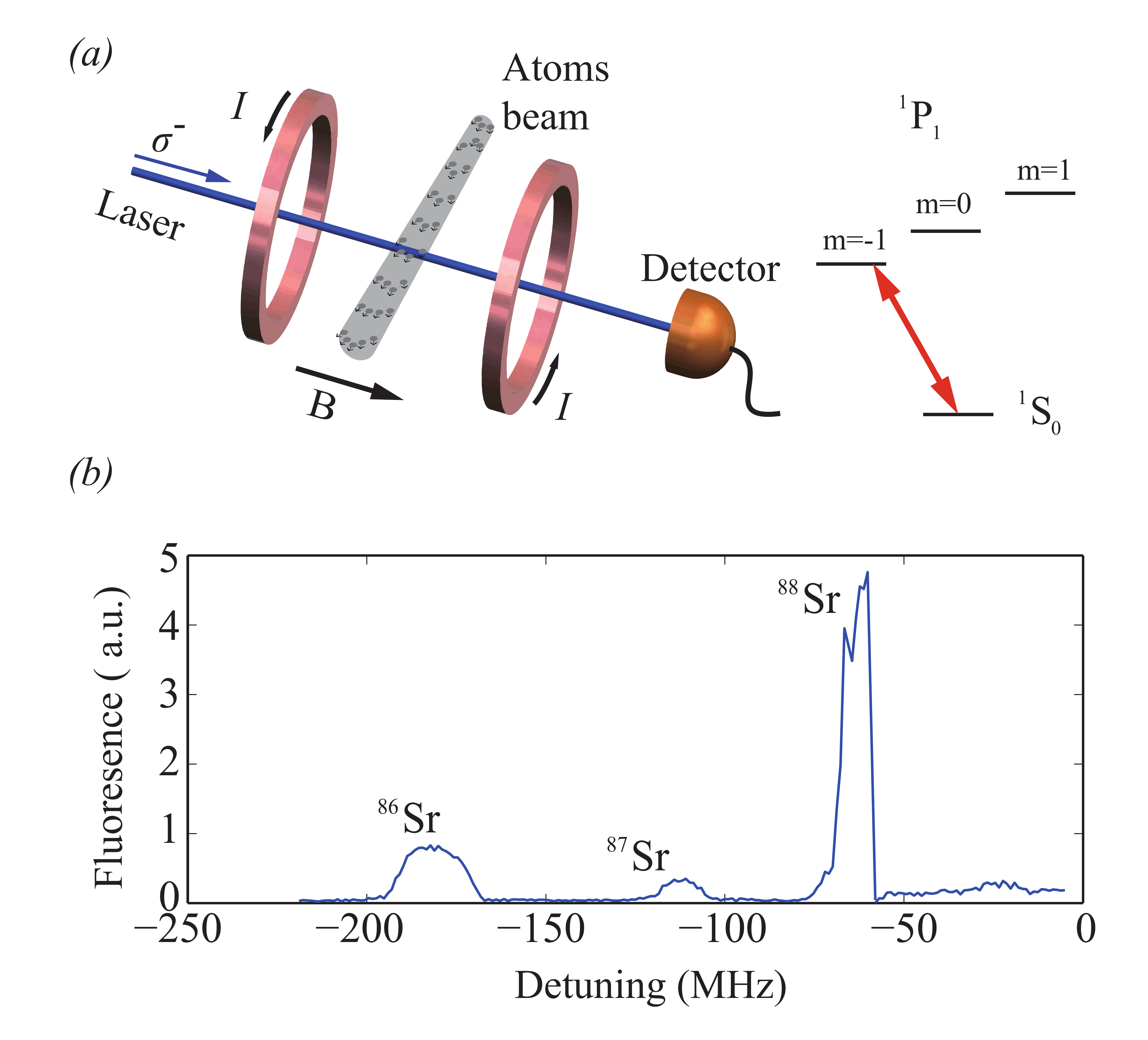}
\caption{(a) Schematic representation of the absorption spectroscopy system used to lock the frequency of the \unit{461}{nm} laser system. A pair of Helmholtz coils generates a bias magnetic field that shifts the atomic transition. (b) The MOT fluorescence spectrum for three Sr isotopes. The origin of the $x$-axis corresponds to the $^{88}\textrm{Sr}$ blue resonance. The frequency scan is performed by changing the value of the bias field.}
\label{Isotopes}
\end{figure}


\subsection{Red laser}
\label{sub_sec_red_laser}

The optical system of the red laser is sketched in Fig.~\ref{RedLaser}. We use a master laser made up of an ECDL of \unit{10}{cm} in length. The output coupler is a diffraction grating with \unit{1800}{lines/mm}. The facet of the intra-cavity \unit{10}{mW} diode laser is anti-reflection coated. Two isolators with an isolation of \unit{60}{dB} and \unit{30}{dB} are placed at the output of the ECDL to avoid any optical feedback. The laser is frequency locked on the reflection signal of an ultra low expansion (ULE) and high finesse ($F=3000$) Fabry-P\'erot (FP) cavity using the \emph{Pound-Drever-Hall} technique. The frequency of the master laser is modulated using an electro-optic modulator (EOM) at \unit{20}{MHz}. We generate the error signal by demodulating at the same frequency. The error signal is sent to the PID controllers which generate feedback to the current controller and the PZT of the master laser for fast and slow corrections respectively. The unitary gain of the feedback loop is at \unit{1}{MHz}. Under this condition, the frequency linewidth is below \unit{1}{kHz} over one second with respect to the FP cavity~\cite{phd_bidel}. The FP cavity is isolated from environmental perturbations by a vacuum chamber. The vacuum chamber is pumped below \unit{10^{-8}}{mbar}, and attached to a double-layer temperature stabilized chamber at 30$^\circ$C. The long-term drift of the FP cavity is below \unit{10}{Hz/s}.

\begin{figure}[h]
\includegraphics[width=0.48\textwidth]{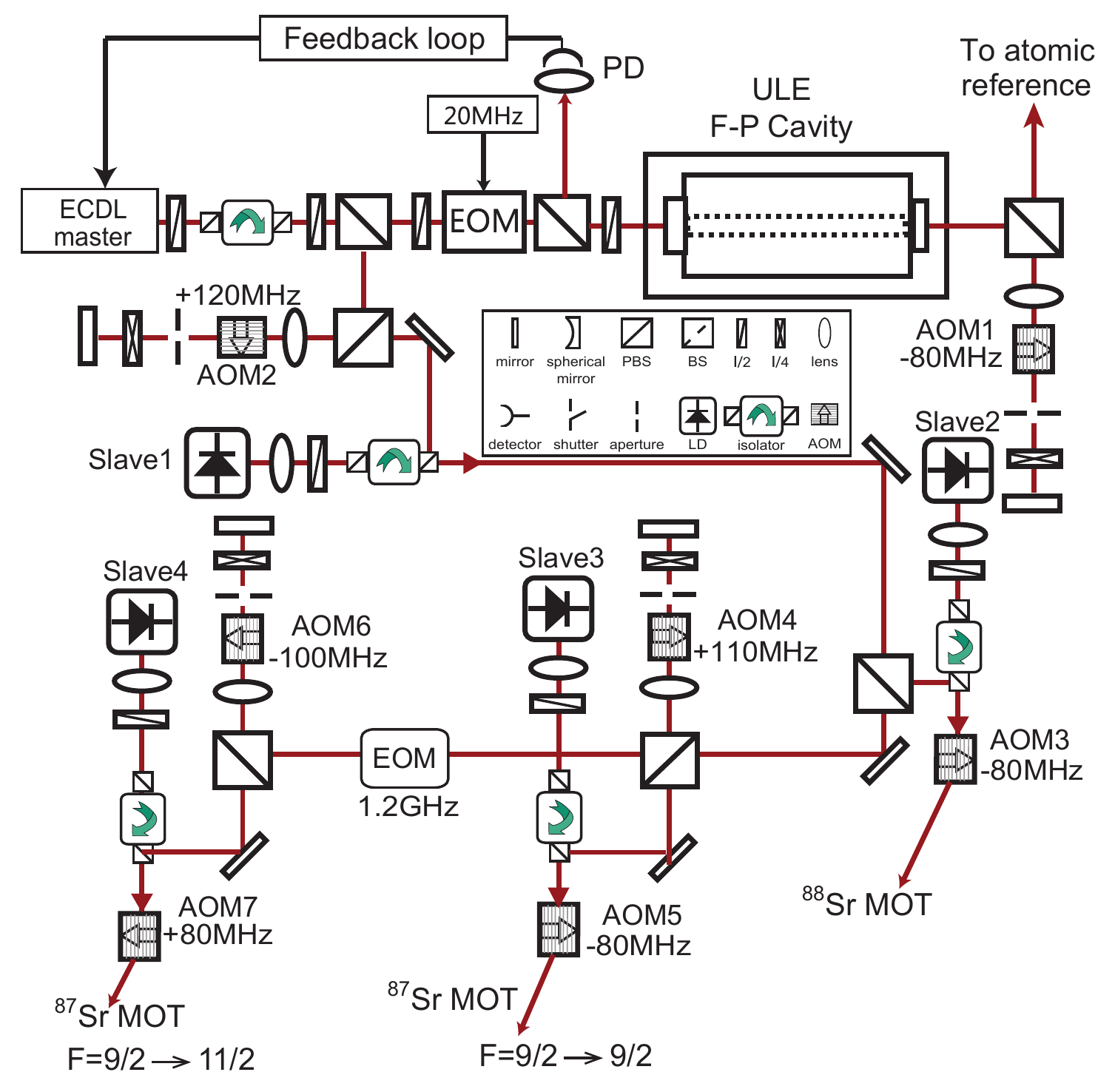}
\caption{Optical setup of the \unit{689}{nm} laser system.}
\label{RedLaser}
\end{figure}


The drift of the FP cavity is compensated using saturated fluorescence spectroscopy on the $^1\snS_0\rightarrow\,^3\snP_1$ transition of $^{88}$Sr. The resonance frequency of the $^{88}$Sr is around \unit{160}{MHz} below the FP cavity resonance. We use an AOM, referred to as AOM1 in Fig.~\ref{RedLaser}, in double pass configuration, and driven at an RF frequency of \unit{80}{MHz} to bring the laser beam transmitted through the FP cavity into resonance with the atomic transition. We apply a bias magnetic field of \unit{10}{G} to split the magnetic sublevels ($m$ = 0, $\pm$1) beyond the Doppler broadening of \unit{9.5}{MHz} that arises from the residual transverse velocity of the atomic beam. We perform saturated fluorescence spectroscopy on the Zeeman shift free $^1\snS_0,m=0\rightarrow\,^3\snP_1,m=0$ transition [Fig \ref{689Sptr}(a)]. The linewidth of the saturated fluorescence spectrum is \unit{140}{kHz} which is limited by the mean transit time of the atoms in the beam and intensity broadening [Fig \ref{689Sptr}(b)]. We use a \unit{150}{\micro W} probe with a beam waist of \unit{10}{mm}. The probe is retro-reflected using a cat's eye setup to minimize the beam misalignment which can induce broadening and shifting of the saturated spectroscopy dip.  We collect the fluorescence signal of the atoms on a femtowatt photodetector (\emph{Thorlabs PDF10A/M}). To increase the fluorescence signal, we also collimate and retro-reflect the fluorescence emission in the opposite direction, as shown in Fig.~\ref{689Sptr}(a). The signal is digitalized and sent to a PC for analysis. The error signal is generated by taking the difference of the photodetector signal while the frequency of the AOM1 is changed back and forth with a \unit{140}{kHz} frequency step. We repeat this measurement periodically at a rate of \unit{1}{s}. The error signal is sent to a numerical PID controller before being interpreted in terms of frequency offset. Then, it is sent to a Direct Digital Synthesizer (DDS) which controls the RF frequency of AOM1. The frequency of the AOM1 is then broadcasted through a local-area-network system and used, in particular, by another DDS to control the RF frequency of AOM2. The latter is located at the output of the master laser. The accuracy of the frequency stabilization chain is tested on the cold cloud of atoms by absorption spectroscopy on the $^1\snS_0,m=0\rightarrow\,^3\snP_1,m=0$ transition. We find a residual frequency bias of \unit{5}{kHz} and a daily variation of \unit{\pm 5}{kHz}.

The master laser beam seeds a slave laser, named slave1, after being double passed through the AOM2. The output of slave1 is further divided into three parts. The first part seeds slave2, which is used for the $^{88}$Sr MOT. The second part of slave1 seeds slave3 which addresses the $F=9/2\rightarrow F=9/2$ transition used for the $^{87}$Sr MOT stirring beam. The third part of slave1 passes through an EOM at a fixed frequency of \unit{1.2}{GHz}, generating mainly the $\pm 1$ lateral orders. We use this beam with multiple frequency lines to seed slave4. We tune the parameters (temperature and current) of slave4 to selectively lock on to the -1 order of the frequency spectrum~\cite{Pramod}. This laser beam is used for the $F=9/2\rightarrow F=11/2$ MOT transition of $^{87}$Sr.

\begin{figure}[h]
\includegraphics[width=0.48\textwidth]{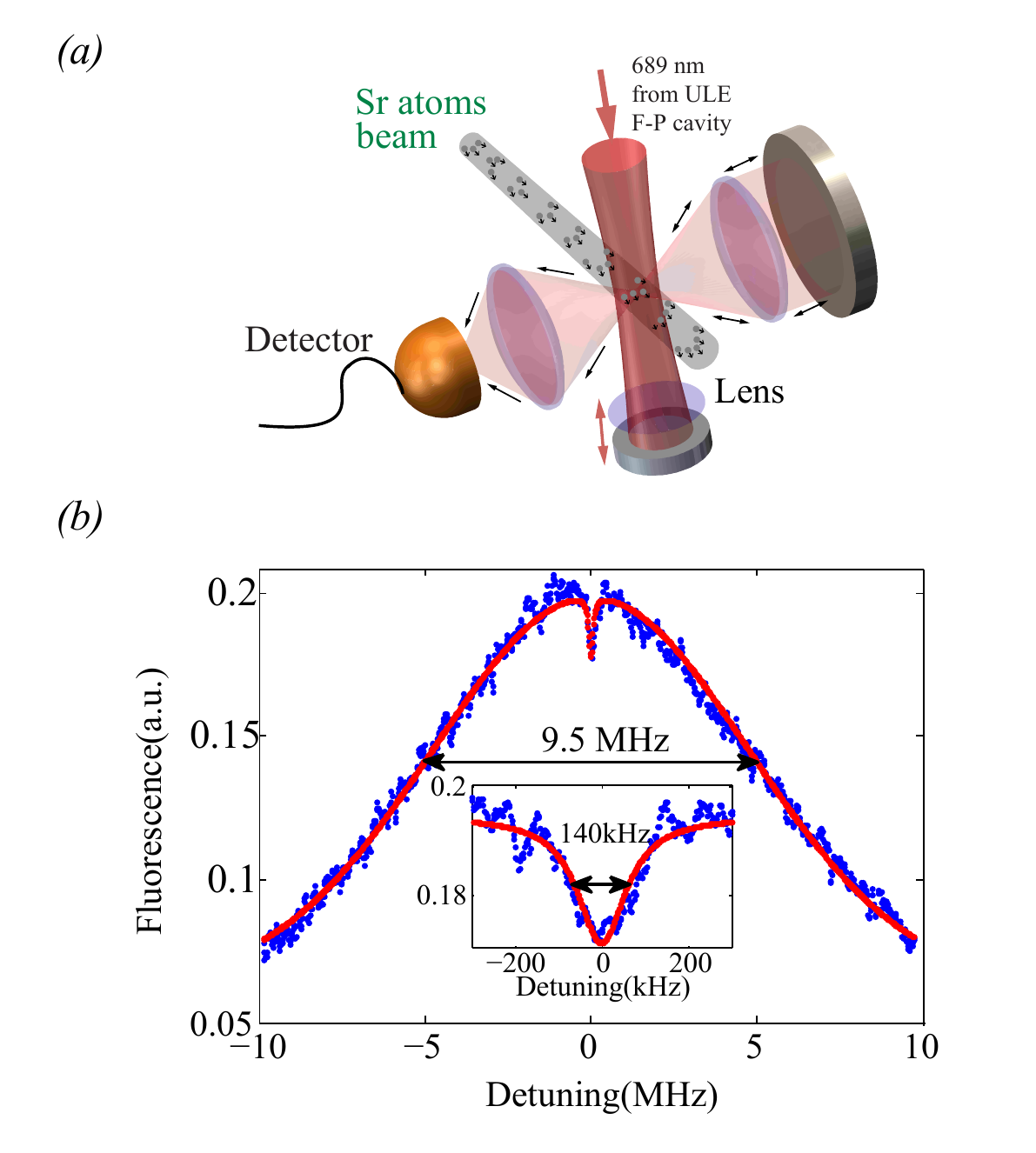}
\caption{(a) Optical setup used for the saturated fluorescence spectroscopy on the $^1\snS_0\rightarrow^3\snP_1$ transition of $^{88}$Sr. (b) Typical saturated fluorescence spectroscopy signal. The blue curve is the experimental spectrum whereas the red curve is a fit of the experimental signal summing a broad gaussian and a narrow lorentzian shape.}
\label{689Sptr}
\end{figure}


\subsection{Repumper lasers}
\label{sub_sec_repumper_laser}

Due to the presence of the low lying energy $^1\snD_2$ state, the blue cooling transition is not completely closed. This leads to the shelving of atoms in the $^3\snP_2$ state (see Fig.~\ref{Srtransitionlevel}). To optically repump the $^3\snP_2$ population to the ground state, we drive the $^3\snP_2\rightarrow\, ^3\snS_1$ transition at \unit{707}{nm}. Since some atoms can be repumped to the long lived $^3\snP_0$ state as well, we also drive the $^3\snP_0\rightarrow\, ^3\snS_1$ transition at \unit{679}{nm}. The \unit{679}{nm} and \unit{707}{nm} lasers are home-made ECDLs. The frequencies of these lasers are monitored and locked by digital feedback on a wavelength meter (HighFinesse, WS/7). The accuracy in frequency of the wavelength meter is \unit{40}{MHz}. It is sufficient since the efficiency of repumping seems to be not affected over a frequency range of \unit{\sim 100}{MHz} across the resonance, most likely because of the spatial inhomogeneities in the transition frequency induced by the MOT magnetic field gradient.

\begin{figure}[h]
\includegraphics[width=0.48\textwidth]{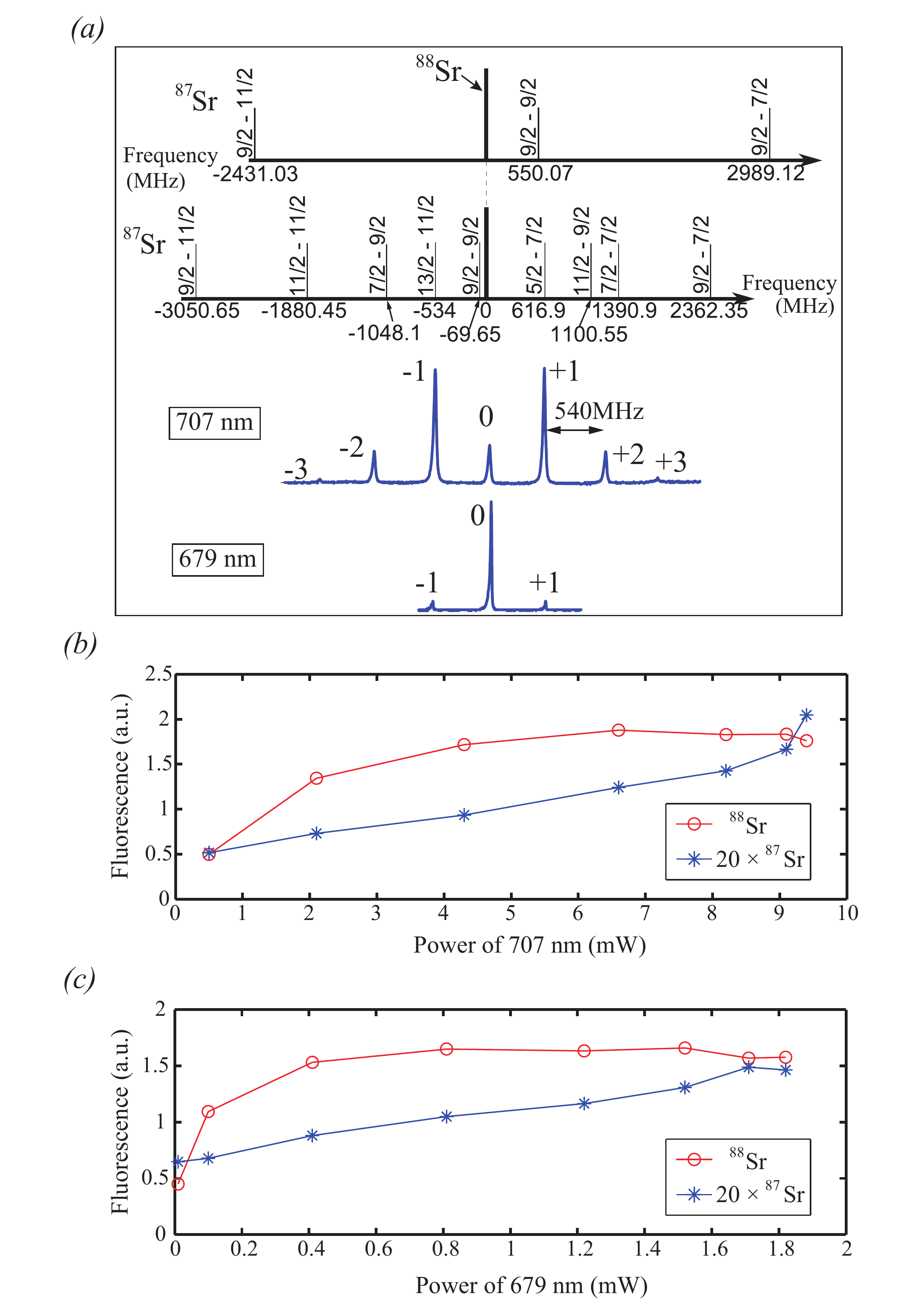}
\caption{Repumping scheme for $^{88}$Sr and $^{87}$Sr. (a) The black solid lines are the different transition for $^{87}$Sr and $^{88}$Sr. The blue curves are the laser spectra after passing through the EOM, modulated at \unit{540}{MHz}. Those spectra are revealed using a low finesse ($F=100$) FP frequency analyzer. The numbers, ranging from $-3$ to $+3$ at the vicinity of the peaks, indicate the lateral bands order of the EOM. (b) Fluorescence signal of the $^{88}\textrm{Sr}$ and $^{87}\textrm{Sr}$ MOT as a function of the \unit{707}{nm} repumper optical power. The \unit{679}{nm} laser is maintained at its maximum power. (c) Fluorescence signal of the $^{88}\textrm{Sr}$ and $^{87}\textrm{Sr}$ MOT as a function of the \unit{679}{nm} repumper optical power. The \unit{707}{nm} laser is maintained at its maximum power.}
\label{Repumper}
\end{figure}


The bosonic case is rather simple, since only two transitions have to be considered. For $^{87}$Sr, the situation is more complicated because of the large hyperfine structure of the $^3\snP_2$ and $^3\snS_1$  levels. The transitions $^3\snP_2\rightarrow\, ^3\snS_1$ and $^3\snP_0\rightarrow\, ^3\snS_1$ are shown in Fig.~\ref{Repumper}(a). In order to pump the population in the $^3\snP_0$ level, we need only one transition as there are no hyperfine levels in this state. \unit{679}{nm} repumper laser addresses the $F=9/2 \rightarrow F=9/2$ transition, which is \unit{550}{MHz} away from the $^{88}$Sr transition. An efficient repumping scheme at \unit{707}{nm} is more challenging, as we need to repump five hyperfine levels, namely $F = 5/2, 7/2, 9/2, 11/2, \textrm{and}\, 13/2$. We choose to address the transitions $F=5/2 \rightarrow F=7/2$, $F=7/2 \rightarrow F=9/2$, $F=9/2 \rightarrow F=9/2$, $F=11/2 \rightarrow F=9/2$, and $F=13/2 \rightarrow F=11/2$, because these transitions are closer to $^{88}$Sr repumping line and they are almost equally separated. We generate a multiline laser using an EOM driven at \unit{540}{MHz} for the \unit{707}{nm} and \unit{679}{nm} laser beams. Both laser beams are mixed using a polarizing beam splitter, thus they have orthogonal polarization. The polarization of the \unit{707}{nm} laser is chosen such that the efficiency of the EOM is optimum. The modulation index of the EOM is adjusted to $1.8$ to get substantial power into the central band and in the $\pm 1$ and $\pm 2$ lateral bands [see Fig.~\ref{Repumper}(a)]. The efficiency of the EOM for the \unit{679}{nm} laser beam is much weaker but it is sufficient to repump the atoms from the $^3\snP_0$ state. The fluorescence signal from the cold cloud in the blue MOT is shown in Fig.~\ref{Repumper}(b,c)} as a function of the \unit{679}{nm} and the \unit{707}{nm} laser power. With this scheme, we are able to repump $^{88}$Sr and $^{87}$Sr without changing the frequency lock points of the lasers.

\section{Conclusion}
\label{sec_conclusion}

We described a new experimental apparatus that produces ultracold gases of $^{88}$Sr and $^{87}$Sr. The system is characterized by a high loading rate of the blue MOT done on the dipole allowed transition. This loading is performed under high vacuum environment. We achieve this by introducing an atomic beam deflector (2D MOT) after the Zeeman slower. Using the 2D MOT, we selectively deflect the longitudinally decelerated atomic beam into the science chamber. We show that the 2D MOT efficiency is limited by the decay into the $^1\snD_2$ long-lived state. The maximum loading rate measured in the MOT is \unit{6\times 10^9}{s^{-1}}. A differential pumping tube is place between the 2D MOT and the science chamber to reduce the background pressure at the MOT location. The lifetime of the atoms in the magnetic trap is around \unit{54}{s}. An extra optical axis is opened up along the atomic beam direction after this deflection. This can be useful, for example, to further manipulate the beam towards a continuous atom laser.

The frequency lock point of the \unit{461}{nm} laser system can be tuned by shifting the atomic transition with Zeeman effect. This technique allows us to easily tune the laser on the desired isotope. The loss mechanism induced by spontaneous decay of the excited state of the cooling transition $^1P_1$ to the underlying long live state $^1D_2$ are properly shelved using a simple multi-line \unit{707}{nm} laser. Finally, we show that a transfer efficiency of 50$\%$ from the blue MOT to the narrow intercombination line red MOT is achieved.

\section{Acknowledgement}
This work was supported by the CQT/MoE funding Grant No. R-710-002-016-271.
%
\bibliographystyle{epj}
\bibliography{HF}%
%
%

\end{document}